# DCT based Fusion of Variable Exposure Images for HDRI

Vivek Ramakrishnan[1] and D. J. Pete[2]

[1]Research Scholar, Department of Electronics Engineering,
Datta Meghe College of Engineering, Sector-3, Airoli,
Navi Mumbai - 400708, India
[2]Professor and Head, Department of Electronics Engineering,
Datta Meghe College of Engineering, Sector-3, Airoli,
Navi Mumbai - 400708, India

*ABSTRACT*

*Combining images with different exposure settings are of prime importance in the field of computational photography. Both transform domain approach and filtering based approaches are possible for fusing multiple exposure images, to obtain the well-exposed image. We propose a Discrete Cosine Trans- form (DCT-based) approach for fusing multiple exposure images. The input image stack is processed in the transform domain by an averaging operation and the inverse transform is performed on the averaged image obtained to generate the fusion of multiple exposure image. The experimental observation leads us to the conjecture that the obtained DCT coefficients are indicators of parameters to measure well-exposedness, contrast and saturation as specified in the traditional exposure fusion based approach and the averaging performed indicates equal weights assigned to the DCT coefficients in this non- parametric and non pyramidal approach to fuse the multiple exposure stack.*

## 1. INTRODUCTION

The classical approach [13] to fuse multiple-exposure image set involves Laplacian pyramid and finding the characteristic parameters like Saturation, Contrast and Well- exposedness and fusing them using the Gaussian weighting scheme. Later approaches like fusing multiple exposure images with like the ones listed in [2], and transform do- main approaches [reference] started to evolve. Pixel based representation of images are possible on paper and traditional display devices. For colour image displays we assign three channels R,G and B channel and assign 8 bit per pixel for each channel. So for every colour channel 256 possible values are formed. Figure 1 depicts the variable exposure stack of the Office images. Figure 2 depicts the generated HDR image. Figure 3 shows the tone-mapped LDR image.

Various HDR formats such as the ones stated below are listed in [39]

1. OPENEXR FORMAT (or EXtended Range format or '.exr' format.).

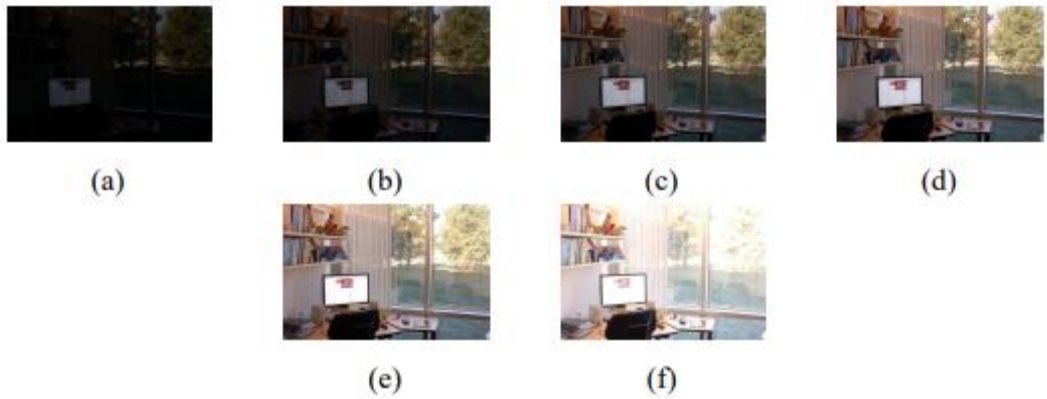

Fig. 1. Set of Six Multiple Exposure Office Images.

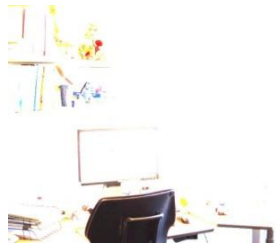

Fig. 2. HDR image of Office

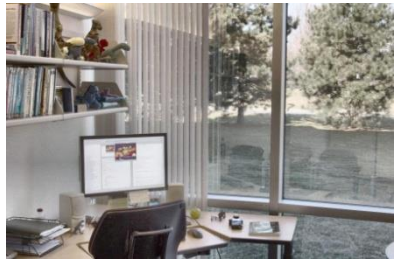

Fig. 3. Tone-mapped image of HDR image of Office

2. Other encodings viz. scRGB48, ScRGB-nl, ScYCC-nl
3. TIFF float and the LOGLUV format.
4. HDR format (.hdr, .pic format), originally known as the Radiance picture format and Lossy HDR encodings.

In this work we perform fusion of the transform domain coefficients using a fusion rule. The multiple exposure stack is first transformed into the DCT domain and then an averaging operation is performed of the transform coefficients. This averaged output is then inverse transformed to obtain the fused output. The block variance is allowed to have a statistical distribution, and the DCT coefficients follow a Laplacian nature.[40] this then implies that we exploit the Laplacian nature of the coefficients which follows similar nature of the Laplacian distribution or Laplacian pyramidal decomposition in the spatial domain on which we perform in traditional exposure fusion.

## 2. CHALLENGES INVOLVED IN HDR IMAGE CAPTURE

Traditional cameras are unable to measure the radiance distribution of the scene and so it becomes unable to capture the full spectral content and dynamic range of the scene. There are inherent limitations in most digital image sensors so as to not make it possible to capture the entire dynamic range of a scene in a single exposure setting. A standard camera records multiple exposures with the right software. Selection of the right pixel to be combined to obtain the fused image is another challenging task.

Inherent assumptions include

1. Capturing device is perfectly linear
2. Each pixel in every exposure needs to be brought into the same irradiance domain
3. Corresponding pixels are averaged across the exposures.
4. Over and under exposed pixels must be excluded

Under practical considerations

1. "Cameras" are not perfectly linear.
2. Registration of various objects in the scene due to camera shake.
3. Moving objects and objects thus the following needs to be considered.
4. Understanding the nature and plotting of the Camera Response CurveFunction (CRF)
5. Involving the computations of the irradiance values.
6. Image registration techniques.
7. Ghost and Lens flare removal.

The further section discusses in detail the classical HDR imaging technique, by Debevec et. al. [2]. Then we describe the classical exposure fusion technique in the spatial domain involving pyramidal decomposition and Gaussian weighting by Mertens et. al. [1] further ahead evolving to the transform domain approaches and to the current work based on a non-pyramidal and uniform weighting approach based on DCT.

## 3. CRF BASED HDRI

The photochemical and the electronic reciprocity of the imaging system is exploited by the algorithm. The Hurter-Driffield characteristic curve summarizes the response of the film to variations in exposure. The curve gives a plot of the optical density D of the processed film versus the logarithm of its exposure setting. Exposure setting is obtained as the product of the irradiance E and exposure-value or exposure-time. We obtain a quantity Z which is function of the exposure setting X of a particular pixel. Consider a function f which is a conglomeration of characteristic function of the film and the non-linearities introduced by other processing steps. Initially we should recover this function f, once we have f we can recover exposure at each pixel as

$$f^{-1}(Z) \qquad (1)$$

The function f is a monotonically increasing function so $f^{-1}(Z)$ is definable. If we know the exposure time δ(t), the irradiance I is obtained as

$$I = Z/(\delta(t)) \qquad (2)$$

The irradiance I is proportional to the radiance L of the scene. So consider a set of exposure times $\delta(t_j)$ for a static scene with constant lighting the film reciprocity equation can be written as

$$Z_{ij} = f(I_i * \delta(t_j)) \qquad (3)$$

Assuming monotonicity of f we find that f is an invertible function so we can write equation (previous) as

$$f^{-1}(Z_{ij}) = I_i * \delta(t_j) \qquad (4)$$

Taking natural logarithms on both the sides we get

$$ln(f^{-1}(Z_{ij})) = ln(I_i) + ln(\delta(t_j)) \qquad (5)$$

Considering $h=f^{-1}(Z)$ we obtain the final equation for the characteristic curve as follows

$$h(Z_{ij}) == ln(I_i) + ln(\delta(t_j)) \qquad (6)$$

The curve representing h is called as characteristic curve which is generalized and is independent of the input image set. Minimum two photographs with varying exposure settings are required to calculate the characteristic curve. Finally the irradiance is obtained as

$$ln(I_i) = \frac{(\sum_{n=1}^{P} w(Z_{ij}) * (h((Z_{ij})) - ln(\delta(t_j))))}{(\sum_{n=1}^{P} w(Z_{ij}))} \qquad (7)$$

Thus we observe that the CRF based HDRI is computationally intensive and depends on log manipulations.

## 4. MERTEN'S CLASSICAL EXPOSURE FUSION APPROACH IN HDR

A camera of limited dynamic range is able to capture a Output referred image rather than a Scene referred image which is possible through HDR. This approach works basically in the spatial domain. Multiple exposure image set is obtained and each image gets treated by the estimated inverse camera response function (CRF). Weighted pixel blending is carried out to obtain the exposure fused image [1]. Sometimes a Multi Resolution Analysis (MRA) based blending is carried out which being a pyramidal de- composition approach gives rise to visible seams and blur. There also arises a need to go for higher order pyramidal decomposition.

## 5. COMPARISON BETWEEN HDR AND LDR ENCODINGS

A scene referred to an output transformation is a one-way irreversible transformation. This transformation is called by a process called as tone mapping. A typical tone map- ping operation is a transformation in which the scene referred pixel is transformed into an output referred pixel. In the HDR image encoding scene referred Ness is given more prominence than an output referred Ness. Scene referred HDR encoding can be always converted to an output referred format but vice-versa is not true. So its a one-way trans formation.

## 6. HDR IMAGERY APPLICATIONS

The following are a list of HDR Imagery applications

1. Physically based rendering (Global Illumination)
2. Remote sensing:
3. Digital Photography:
4. Image Editing:
5. Digital Cinema (and Video):
6. Virtual Reality (VR):
7. Movie film stock.

## 7. RELATED WORK

Debevec et. al. [8] developed a technique in which different expsoure images are combined into a single radiance map such that the pixel values are in proportion to the radiance values. Bogoni and Hansen [24] stated an in which the Luma and the Chroma components were decomposed into Laplacian pyramid and Gaussian pyramid respectively. A fusion based on Illumination estimation was suggested by [25], Vonikakis et. al. The problem of motion blur in longer exposure images was addressed by Tico et. al. [26]. In the method developed by Jinho and Okuda [27] they designed weighting functions that not dependant on exposure areas. Shen [28] used Generalized random walks to arrive at an optimal solution. Li and Kang [29] suggested a weighted sum fusion approach. Independent Component Analysis (ICA) was used by Mitianoudis and Stathaki [30] to propose their fusion scheme. Fusion involving weighted averaging of the input image stack was proposed in ([5], [6],[7]). Fusion approaches based on the irradiance values, and not the intensity values of the pixels was developed in ( [8], [9]). Transform based approach are discussed in ([38], [4]). Goshtas by developed a block based fusion approach [10]. Compositing without Matte was developed in [12] and the classical exposure fusion approach in [13] are two other important techniques of compositing. Bi lateral filter based method is suggested in [1]. A Bayesian approach for Matte generation is suggested in [11]. Approaches related to Markov random field [14], boundary location information [15], Poisson Solver based [16] and an optimization approach in [17] are other fully developed techniques. Making use of the fore(back)ground statistics an approach was developed in [18], using motion cues in [19] and through defocus cue in [20]. A generalistic $\alpha$ blending approach is stated in 21].

## 8. SOME POPULAR HDR IMAGING TECHNIQUES USED FOR COMPARISON IN THIS WORK

We now give a brief account of some state of the art HDR imaging techniques which are popular and can be used for comparison with this technique. Ashikmin et. al.[31]

developed a technique which approach which maps the HDR image input to the luminance values possible to displayed by the display device, Banterlee et. al. [32] presented a luminance zone based approach where every zone is processed by a Tone Mapping Operator (TMO) compatible for each zone. Raman et. al. [1] developed a bilateral filter based compositing. Bruce et. al. [33] developed a fusion approach based on relative entropy. Lischinski et. al. A method based on interactive local adjustment of tonal values was proposed by [34]. Tan et. al. [35] provides a logarithmic tone mapping algorithm. Mertens et. al. [13] present a parametric classical multiple exposure fusion approach. Reinhard et. al. [36] present a traditional tone mapping methodology. In addition we have transform domain approaches [4] and edge preserving Savitsky Golay filtering based approach [37].

## 9. THE DISCRETE COSINE TRANSFORM (DCT)

The Discrete Cosine Transform used mainly in JPEG compression is a block transform. The basic formulation of the DCT was done by Ahmed et al. [46]. DCT exhibits unitary property and is similar to the KarhaunenLoe`ve Transform (KLT). It has got good energy compaction efficiency and rate distortion function.. In the KLT [47] we observe decorrelated transform coefficients and the signal energy is compacted in the first fewest sub-bands. The N x N cosine transform matrix V = v (k, n), also known as the Discrete Cosine Transform (DCT) which is defined as follows for k = 0 and $0 \leq n \leq (N − 1)$

$$v(k,n) = \frac{1}{\sqrt{N}} \tag{8}$$

and for $0 \leq k \leq (N − 1)$, $0 \leq n \leq (N − 1)$

$$v(k,n) = \frac{\sqrt{2}}{\sqrt{N}} * \cos \frac{\pi(2n+1)k}{2N} \tag{9}$$

The properties of DCT are

1. DCT is not the real part of the unitary Discrete Fourier Transform (DFT)
2. DCT is real orthogonal
3. DCT is a fast transform.
4. Excellent energy compaction efficiency is seen in the DCT for highly correlated data.
5. The basis vectors of DCT are eigenvectors of the symmetric tridiagonal matrix.
6. DCT is very close to the KLT.

## 10. REGARDING DISTRIBUTION OF 2-D BLOCK DCT COEFFICIENTS

Pratt (1978) [41] carried out the earlier work on DCT Coefficients, it was conjectured that the DC coefficients are Rayleigh distributed assuming no level shift and the AC coefficients are Gaussian distributed according to the central limit theorem considering each pixel to be statistically independent from another. Tescher (1979) [42] and Mu- rakami, Hatori and Yamamoto (1982) [43] indicated that the AC coefficients are Laplacian distributed, and the DC coefficients are Gaussian distributed. Reininger and Gibson [44] also asserted that DC coefficient follows a Gaussian PDF, and AC coefficients fol- low a Laplacian PDF. Reininger and Gibson performed the Kolmogorov-Smirnov (KS) test [45] on a 256x256 8-bit PCM grayscale images over multiple modes, and deter- mined the best fit for the different PMFs. Observing the results of the (KS) tests [44] it is concluded that the AC block DCT coefficients follow a Laplacian distribution.

## 11. MOTIVATION

The drawbacks of fusion in the spatial domain include

1. Brightness changes are vast around the edges.
2. Low pass filtered version is subtracted from the original to get the Laplacian pyramid based decomposition.
3. Pixels with higher frequency that is higher rate of change of illumination are retained.
4. We get reduced number pixels by sub-sampling the edges.

5. Sub-sampling gives disadvantage of approximating the non-edge pixels with the highly varying brightness edge pixels.
6. The weight maps also depict severe variations.
7. There is degradation in the contrast and there are low frequency brightness change in the pixels which are not edge pixels.

Few of the initial steps which are performed to get a low pass filtered version and subtracting it to get the Laplacian decomposition are achieved by the one step DCT transform itself. The finding of the pixel coefficients for compositing the images is simplified due to uniform weights being assigned to each pixel and the averaging operation. We achieve a seamless blending and the results are comparable with standard exposure fusion based approach and NSST based approach.

## 12. OUR METHOD

The steps involved in our algorithm are as follows

1. Perform the DCT transform of all the images in the image stack.
2. The DCT Coefficients of all these images are averaged out to obtain the resultant final image in the transform domain.
3. The inverse DCT transform of the resultant image in the transform domain is performed to obtain the fused result.

Thus we can note that all the pixels have equal prominence so equal weights are assigned to all the pixels which are captured in the transform domain. The transform coefficients are empowered enough to give prominence in their luma value and chroma values according to the exposure settings.

## 13. RESULTS AND DISCUSSION

The DCT-based exposure fusion outputs are compared with the techniques described in ([1],[4],[13],[31],[32],[33],[34],[35],[36] and [37] ). The PSNR and the SSIM scores are improved. There is a decrease in error i.e. the IMMSE scores. High Q Scores are obtained using the VDP metric [23]. For the following set of images, shown in figures 5, 7, 9, 11 kindly refer to the table 1 shown below for knowing which of the sub-figure was generated using what approach.

Table 1. List of the sub figures indexes, showing results for images generated using various approaches used for comparison in this work.

| Sub figure index | Results using method described in |
|---|---|
| a | S G Filter based [37] |
| b | Ashikmin et. al. [31] |
| c | Banterlee et. al. [32] |
| d | Shanmuga et. al. [1] |
| e | Bruce et. al. [33] |
| f | Lischinski et. al. [34] |
| g | Logarithmic et. al. [35] |
| h | Mertens et. al. [13] |
| i | Reinhard et. al. [36] |
| j | NSST Based [4] |
| k | Our Method in this paper |

The PSNR, SSIM, IMMSE scores and the Q factor score [23] are tabulated in the table 2.

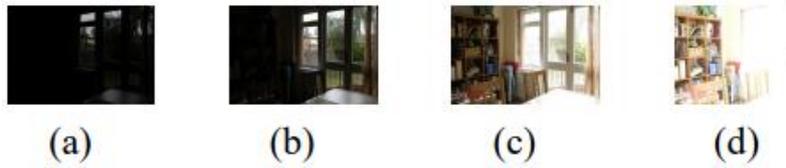

Fig. 4. Variable Exposure House Images.

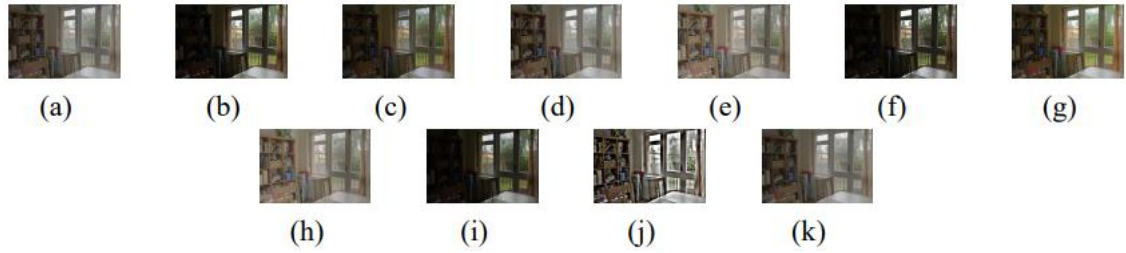

Fig. 5. Results generated for House images with approaches mentioned in table 1.

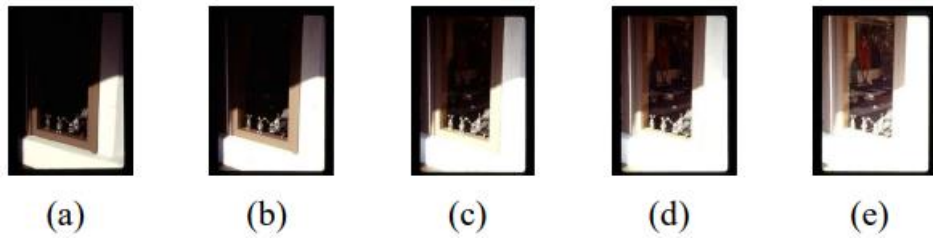

Fig. 6. Variable exposure Window Images.

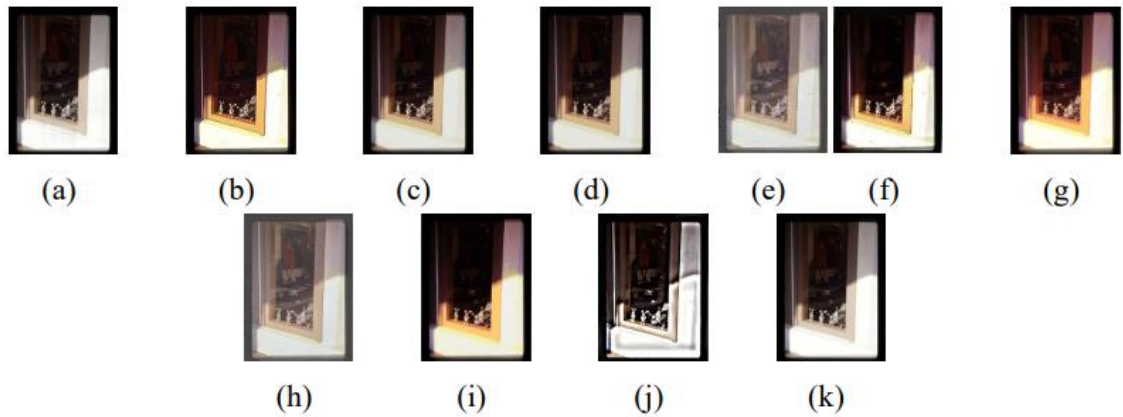

Figure 7. Results generated for Windows images with approaches mentioned in table 1.

Computer Science & Information Technology (CS & IT) 197

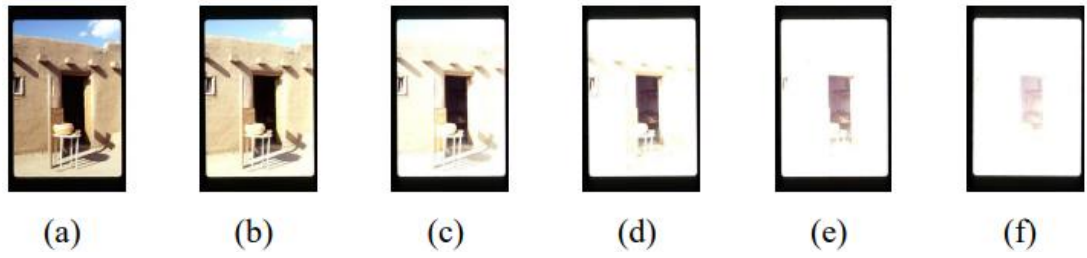

Figure 8. Variable exposure Door Image set.

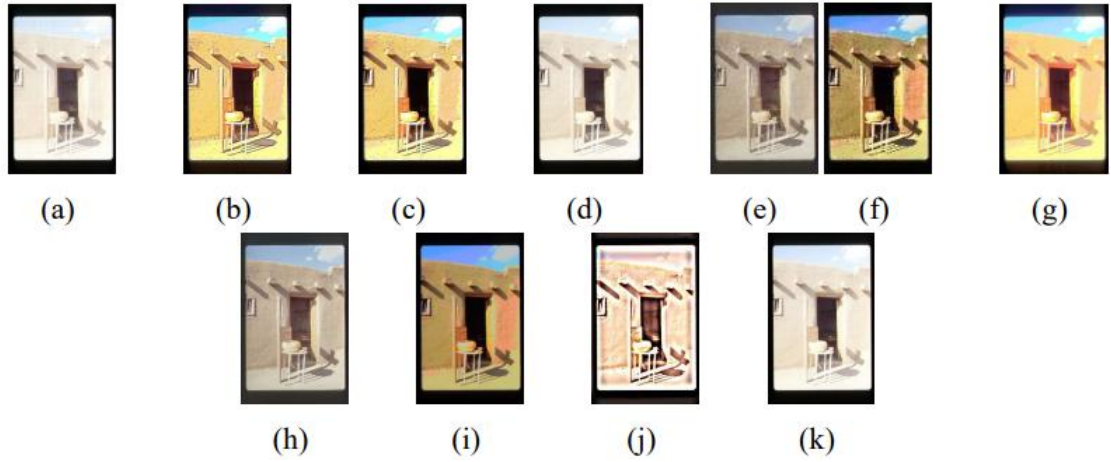

Figure 9. Results generated for Door images with approaches mentioned in table 1.

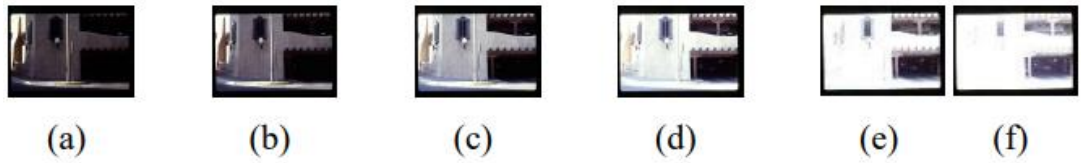

Figure 10. Variable Exposure Garage Image set.

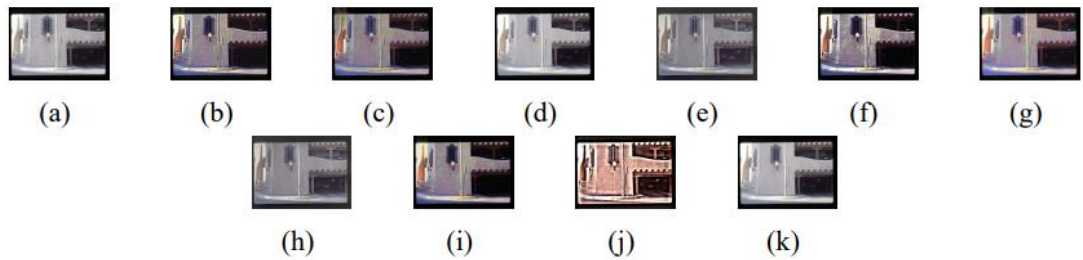

Figure 11. Results generated for Garage images with approaches mentioned in table 1

Table 2. Performance parameters for the DCT-based Exposure Fusion and other state of the art approaches described in ([1],[4],[13],[31],[32],[33],[34],[35],[36]and[37])

| Image | Method | PSNR | IMMSE | SSIM | Q |
|---|---|---|---|---|---|
| HOUSE IMAGE | *Ashikmin*[31] | 66.1274 | 0.0300 | 0.9877 | 98.0525 |
| | *Banterlle*[32] | 70.1584 | 0.0179 | 0.9890 | 98.1235 |
| | *BFbased*[1] | 67.0624 | 0.0261 | 0.9878 | 98.2014 |
| | *Bruce*[33] | 62.8707 | 0.0523 | 0.9839 | 98.1466 |
| | *Lischinski*[34] | 64.7319 | 0.0375 | 0.9865 | 98.0537 |
| | *Logarithmic*[35] | 68.5648 | 0.0214 | 0.9885 | 98.1503 |
| | *Mertens*[13] | 62.8707 | 0.0523 | 0.9839 | 98.466 |
| | *Reinhard*[36] | 63.5942 | 0.0366 | 0.9869 | 98.1023 |
| | *SGFilterBased*[37] | 68.1417 | 0.0226 | 0.9883 | 98.1874 |
| | *NSSTBased*[4] | 63.6711 | 0.0452 | 0.9866 | 98.0366 |
| | *DCTBased(OurMethod)* | 67.7379 | 0.0238 | 0.9881 | 98.1997 |
| DOOR IMAGE | *Ashikmin*[31] | 61.7326 | 0.0649 | 0.9838 | 98.0960 |
| | *Banterlle*[32] | 61.5589 | 0.0672 | 0.9834 | 98.1026 |
| | *BFbased*[1] | 70.6514 | 0.0170 | 0.9892 | 98.2521 |
| | *Bruce*[33] | 61.1196 | 0.0733 | 0.9892 | 98.1909 |
| | *Lischinski*[34] | 58.2465 | 0.1326 | 0.9748 | 98.0905 |
| | *Logarithmic*[35] | 62.2420 | 0.0589 | 0.9842 | 98.169 |
| | *Mertens*[13] | 61.1196 | 0.0733 | 0.9822 | 98.1909 |
| | *Reinhard*[36] | 58.0458 | 0.1384 | 0.9737 | 98.1865 |
| | *SGFilterBased*[37] | 70.89 | 0.0167 | 0.9893 | 98.2389 |
| | *NSSTBased*[4] | 63.8374 | 0.0438 | 0.9864 | 98.1472 |
| | *DCTBased(OurMethod)* | 71.0176 | 0.0165 | 0.9893 | 98.2925 |
| WINDOW IMAGE | *Ashikmin*[31] | 65.4498 | 0.0333 | 0.9875 | 98.0258 |
| | *Banterlle*[32] | 65.4498 | 0.0333 | 0.9875 | 98.0258 |
| | *BFbased*[1] | 67.2379 | 0.0255 | 0.9882 | 98.0855 |
| | *Bruce*[33] | 61.0632 | 0.0745 | 0.9810 | 98.0566 |
| | *Lischinski*[34] | 65.4698 | 0.0332 | 0.9875 | 98.0036 |
| | *Logarithmic*[35] | 66.7944 | 0.0271 | 0.9881 | 98.0285 |
| | *Mertens*[13] | 61.0368 | 0.0745 | 0.9810 | 98.0566 |
| | *Reinhard*[36] | 66.8427 | 0.0719 | 0.9881 | 98.0362 |
| | *SGFilterBased*[37] | 70.6617 | 0.0170 | 0.9892 | 98.1223 |
| | *NSSTBased*[4] | 66.9991 | 0.0263 | 0.9883 | 98.1374 |
| | *DCTBased(OurMethod)* | 71.7756 | 0.0154 | 0.9894 | 98.1227 |
| GARAGE IMAGE | *Ashikmin*[31] | 66.4497 | 0.0285 | 0.9880 | 98.0795 |
| | *Banterlle*[32] | 65.1389 | 0.0351 | 0.9871 | 98.0327 |
| | *BFbased*[1] | 70.0185 | 0.0182 | 0.9890 | 98.2360 |
| | *Bruce*[33] | 63.7948 | 0.0442 | 0.9854 | 98.1752 |
| | *Lischinski*[34] | 67.6633 | 0.0240 | 0.9884 | 98.0466 |
| | *Logarithmic*[35] | 68.7227 | 0.0210 | 0.9888 | 98.0713 |
| | *Mertens*[13] | 63.7948 | 0.0442 | 0.9854 | 98.1752 |
| | *Reinhard*[36] | 67.2374 | 0.0255 | 0.9882 | 98.1057 |
| | *SGFilterBased*[37] | 70.6275 | 0.0171 | 0.9891 | 98.2424 |
| | *NSSTBased*[4] | 63.6227 | 0.0455 | 0.9862 | 98.1145 |
| | *DCTBased(OurMethod)* | 70.8156 | 0.0168 | 0.9891 | 98.2364 |

## 14. CONCLUSIONS

Thus we have developed a method for performing fusion of multiple exposure images using the DCT-based transform domain approach and studied the results. The results favor the use of DCT-based multi-exposure image fusion forthe High Dynamic Range Imaging problem.

**ACKNOWLEDGEMENTS**

Tom Mertens, Frank Van Reeth and Jan Kautz are thanked for the set of multiple exposure images in Fig.(4). The CAVE Computer Vision Laboratory, Columbia University is thanked for the multiple exposure images in the Figures, Fig.(6) , Fig.(8) and Fig.(10).

**REFERENCES**

[1] Raman, S. and S. Chaudhuri. "Bilateral Filter Based Compositing for Variable Exposure Photography." Eurographics (2009).

[2] B. G. Gowri, V. Hariharan, S.Thara, V. Sowmya, S. S. Kumar and K. P. Soman, "2D Image data approximation using Savitzky Golay filter — Smoothing and differencing," 2013 Inter- national Mutli-Conference on Automation, Computing, Communication, Control and Com- pressed Sensing (iMac4s), Kottayam, 2013, pp. 365-371, doi: 10.1109/iMac4s.2013.6526438. [3]. T. Porter and T. Duff. Compositing digital images. In ACM Siggraph Computer Graphics, volume 18, pages 253–259 ACM, 1984.

[4] Ramakrishnan, V., Pete, D.J. Non Subsampled Shearlet Transform Based Fusion of Multiple Exposure Images. SN COMPUTER SCIENCE 1, 326 (2020). https://doi.org/10.1007/s42979- 020-00343-4.

[5] Ron Brinkmann. 1999. The art and science of digital compositing. Morgan Kaufmann Pub- lishers Inc., San Francisco, CA, USA.

[6] Thomas Porter and Tom Duff. 1984. Compositing digital images. SIGGRAPH Comput Graph. 18, 3 (July 1984), 253–259. DOI:https://doi.org/10.1145/964965.808606

[7] J. F. Blinn, "Compositing. 1. Theory," in IEEE Computer Graphics and Applications, vol. 14, no. 5, pp. 83-87, Sept. 1994, doi: 10.1109/38.310740.

[8] Debevec, and J. Malik. Recovering High Dynamic Range Radiance Maps from Photographs P SIGGRAPH 97 (August 1997).

[9] MANN S., PICARD R. W.: On being undigital with digital cameras: extending dynamic range by combining exposed pictures. In In Proc. of IS & T 48th annual conference (1995), pp. 422–428.

[10] GOSHTASBY A.: Fusion of multi-exposure images. Image and Vision Computing 23 (2005), 611–618.

[11] Y. Chuang, B. Curless, D. H. Salesin, and R. Szeliski. A bayesian approach to digital matting. In CVPR, volume 2, pages 264–271, Kauai Marriott, Hawaii, 2001.

[12] RAMAN S., CHAUDHURI S.: A matte-less, variational approach to automatic scene com- positing. In ICCV (2007).

[13] MERTENS T., KAUTZ J., REETH F. V.: Exposure fusion. In Pacific Graphics (2007). [14].Y. Guan, W. Chen, X. Liang, Z. Ding, and Q. Peng. Easy matting - a stroke based approach for continuous image matting. Eurographics, 25(3):567–576, 2006.

[15]. M. Ruzon and C. Tomasi. Alpha estimation in natural images. In CVPR, volume 1, pages 18–25, Hilton Head Island, South Carolina, USA, 2000.

[16]. J. Sun, J. Jia, C. Tang, and H. Shum. Poisson matting. In SIGGRAPH, pages 315–321, Los Angeles, USA, 2004.

[17]. J. Wang and M. F. Cohen. An iterative optimization approach for unified image segmentation and matting. In ICCV, pages 936–943, Beijing, China, 2005.

[18] N. Apostoloff and A. Fitzgibbon. Bayesian video matting using learnt image priors. In CVPR, volume 1, pages 407–414, Washington, DC, USA, 2004.

[19] Y. Chuang, A. Agarwala, B. Curless, D. H. Salesin, and R. Szeliski. Video matting of com- plex scenes. In SIGGRAPH, pages 243–248, San Antonio, USA, 2002.

[20] M. McGuire, W. Matusik, H. Pfister, J. F. Hughes, and F. Durand. Defocus video matting. In SIGGRAPH, pages 567–576, Los Angeles, USA, 2005. A. Smith. Alpha and the history of digital compositing. Microsoft Tech Memo 7, 1995.

[21] 'Savitzky Golay filter for 2D images' - http://research.microsoft.com/enus/um/people/jckrumm/SavGol/SavGol.htm, September 2012. [22].Mantiuk R., Kim K., Rempel A.,Heidrich W. (2011). HDR-VDP-2: A calibrated visual met- ric for visibility and quality predictions in all luminance conditions ACM Trans. Graph.. 30. 40. 10.1145/1964921.1964935.

[23] Bogoni, L.; Hansen, M. Pattern-selective color image fusion. Pattern Recognit. 2001, 34, 1515–1526.
[24] Vonikakis, V.; Bouzos, O.; Andreadis, I. Multi Exposure Image Fusion Based on Illumination Estimation. In Proceedings of the SIPA, Chania, Greece, 22–24 June 2011; pp. 135–142.
[25] Tico, M.; Gelfand, N.; Pulli, K. Motion-Blur-Free Exposure Fusion. In Proceedings of the 2010 IEEE International Conference on Image Processing, Hong Kong, China, 26–29 Septem- ber 2010; pp. 3321–3324.
[26] Jinno, T.; Okuda, M. Multiple Exposure Fusion for High Dynamic Range Image Acquisition. IEEE Trans. Image Process. 2012, 21, 358–365.
[27] Shen, R.; Cheng, I.; Shi, J.; Basu, A. Generalized Random Walks for Fusion of Multi- Exposure Images. IEEE Trans. Image Process. 2011, 20, 3634–3646.
[28] Li, S.; Kang, X. Fast multi-exposure image fusion with median filter and recursive filter. IEEE Trans. Consum. Electron. 2012, 58, 626–632
[29] Mitianoudis, N.; Stathaki, T. Pixel-based and Region-based Image Fusion schemes using ICA bases. Inf. Fusion 2007, 8, 131–142.
[30] Michael Ashikhmin. 2002. A tone mapping algorithm for high contrast images. In Proceed- ings of the 13th Eurographics workshop on Rendering (EGRW '02). Eurographics Association, Goslar, DEU, 145–156.
[31]. Francesco Banterle, Alessandro Artusi, Elena Sikudova, Thomas Bashford-Rogers, Patrick Ledda, Marina Bloj, and Alan Chalmers. 2012. Dynamic range compression by differential zone mapping based on psychophysical experiments. In Proceedings of the ACM Symposium on Applied Perception (SAP '12). Association for Computing Machinery, New York, NY, USA, 39–46. DOI:https://doi.org/10.1145/2338676.2338685
[32] Bruce, N.D. (2014). ExpoBlend: Information preserving exposure blending based on nor-
[33] malized log-domain entropy. Comput. Graph., 39, 12-23.
[34] Lischinski, D., Farbman, Z., Uyttendaele, M., Szeliski, R. (2006). Interactive local adjust- ment of tonal values. ACM Transactions on Graphics (TOG), 25(3), 646-653.
[35] Tan, J., Huang, Y., Wang, K. (2018, July). Logarithmic Tone Mapping Algorithm Based on Block Mapping Fusion. In 2018 International Conference on Audio, Language and Image Processing (ICALIP) (pp. 168-173). IEEE.
[36] Reinhard, E., Stark, M., Shirley, P., Ferwerda, J. (2002, July). Photographic tone reproduction for digital images. In Proceedings of the 29th annual conference on Computer graphics and interactive techniques (pp. 267-276).
[37] Ramakrishnan, Vivek., Pete, D.. (2021). Savitzky–Golay Filtering-Based Fusion of Mul- tiple Exposure Images for High Dynamic Range Imaging. SN Computer Science. 2. 10.1007/s42979-021-00594-9.
[38] Vivek Ramakrishnan "Exposure Fusion in the Non-Sub-sampled Contourlet Domain " Vol. 9 - No. 2 (Feb 2019), International Journal of Engineering Research and Applications (IJERA) ISSN: 2248-9622 , www.ijera.com Last accessed 08 July 2021.
[39] Erik Reinhard, Greg Ward, Sumanta Pattanaik, and Paul Debevec. 2005. High Dynamic Range Imaging: Acquisition, Display, and Image-Based Lighting (The Morgan Kaufmann Se- ries in Computer Graphics). Morgan Kaufmann Publishers Inc., San Francisco, CA, USA. [40].Lam E. Y. , "Analysis of the DCT coefficient distributions for document coding," in IEEE Signal Processing Letters, vol. 11, no. 2, pp. 97-100, Feb. 2004, doi: 10.1109/LSP.2003.821789. [41].Pratt W. K., Digital Image Processing, New York: Wiley-Interscience, 1978, chapter 10. [42].Tescher A.G., "Transform image coding," in Advances in Electronics and Electron Physics. Suppl. 12. New York: Academic, 1979, pp. 113-115.
[43] Murakami H, Hatori Y., and Yamamoto H., "Comparison between DPCM and Hadamard transform coding in the composite coding of the NTSC color TV signal," IEEE Transactions on Communications, vol COM-30, pp. 469-479, Mar. 1982.
[44] Reininger R., and Gibson J. : "Distribution of the two-dimensional DCT coefficients for images," IEEE Transactions on Communications 31 (6) 1983.
[45] Weisstein E. W., "Kolmogorov-Smirnov Test." From Math World–A Wolfram Web Re- source. http://mathworld.wolfram.com/Kolmogorov-SmimovTest.html, last viewed on March 22, 2010.
[46] Ahmed N., Natarajan T., and Rao K. R., "Discrete Cosine Transform," IEEE Transactions on Computers, 90-93, Jan 1974.
[47] Akansu, Ali N. and Haddad, Richard A., Multiresolution Signal Decomposition, Transforms, Subbands, Wavelets, Second Edition, San Diego, CA, Academic Press, 2001.

## AUTHORS

**Aut Mr. Vivek Ramakrishnan**, is a Research scholar in the Electronics engineering department, Datta Meghe College of Engineering, Airoli, Navi-Mumbai, India. He is a researcher in the area of Signal, Image and Video Processing, Computational Photography and Medical Imaging. His current research areas are transform and filtering based approaches for HDRI. He has many research publications to his credit in the said domain.

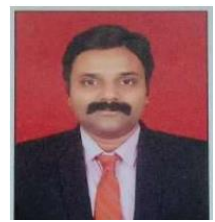

**Dr. D. J. Pete**, is Professor and Head, in the Electronics engineering department, Datta Meghe College of Engineering, Airoli, Navi-Mumbai, India. He has teaching experience of over 24 years His areas of expertise include Communication Engineering, VLSI Reliability and Nano Electronics.

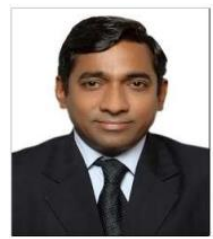